\documentclass[aps,preprint]{revtex4}
\usepackage{amsmath,amssymb,amsthm,mathrsfs,amsfonts,dsfont} 
\usepackage{graphicx}
\usepackage{subfigure}
\usepackage{amsbsy}
\usepackage{bm}
\usepackage{algorithm}
\usepackage{algpseudocode}
\usepackage{algorithmicx}

\def\bx{\mathbf{x}}
\def\by{\mathbf{y}}

\begin{document}

\title{How high dimensional neural dynamics are confined in phase space}
\author{Shishe Wang$^{1}$}
\author{Haiping Huang$^{1,2}$}
\email{huanghp7@mail.sysu.edu.cn}
\affiliation{$^{1}$PMI Lab, School of Physics, Sun Yat-sen University, Guangzhou 510275, People's Republic of China}
\affiliation{$^{2}$Guangdong Provincial Key Laboratory of Magnetoelectric Physics and Devices, Sun Yat-sen University, Guangzhou 510275, People's Republic of China}
\date{\today}

\begin{abstract}
	High dimensional dynamics play a vital role in brain function, ecological systems, and neuro-inspired machine learning. Where and how these dynamics are confined in the phase space remains challenging to solve. Here, we provide an analytic argument that the confinement region is an M-shape when the neural dynamics show a diversity, with two sharp boundaries and a flat low-density region in between. Despite increasing synaptic strengths in a neural circuit, the shape remains qualitatively the same, while the left boundary is continuously pushed away. However,
	in deep chaotic regions, an arch-shaped confinement gradually emerges. Our theory is supported by numerical simulations on finite-sized networks. This analytic theory opens up a geometric route towards addressing fundamental questions about high dimensional non-equilibrium dynamics.
\end{abstract}

 \maketitle

\section{Introduction}
High dimensional neural dynamics support a variety of cognitive functions, such as planning, decision making, and working memory~\cite{Vyas-2020,Ostojic-2024}. Intrinsic structures of neural dynamics receive intensive research interests from diverse fields~\cite{Fiete-2019,Fusi-2023,Helias-2022,Clark-2023,Zou-2023}. The dynamics are computationally modeled by recurrent neural networks~\cite{WCM-1972,Amari-1977,Chaos-1988,Abbott-2009}, where neurons are non-reciprocally connected. The recurrent neural network is also a powerful tool in designing machine learning algorithms~\cite{Eoc-1990,Eoc-2004,Maass-2009}, and was even used to explain consciousness~\cite{PNAS-2022}. This computational model is thus an impactful metaphor for brain dynamics and further computational intelligence.

Mechanistic interpretation of this model started from identifying a chaos transition in a random recurrent connection setting~\cite{Chaos-1988}. This dynamical complexity was then revealed to be coupled with topological complexity, i.e., the emergence of exponentially many unstable fixed points~\cite{PRL-2013}. This finding was recently revised and extended to deep chaotic region~\cite{Helias-2022}. In particular, this recent work demonstrated a radial separation of fixed points and dynamics, which marks an important step toward fully understanding high-dimensional neural dynamics. Another interesting work argued that the linear dimensionality of chaotic attractors increases from the transition point until saturation~\cite{Clark-2023}. These works clarified computational mechanisms based on either the Kac-Rice formula counting the number of fixed points~\cite{Kac,Rice,PRL-2013,Helias-2022} or the dynamical mean-field theory~\cite{Zou-2024,Clark-2023}. 
However, these physics approaches are unable to identify the specific shape of the phase space that confines the neural dynamics, particularly at different levels of dynamics slowness (explained below), and thus the underlying mathematical foundation is still obscure. Therefore, the long-lasting fundamental problem of where and how the high dimensional neural dynamics are confined remains under debate. Solving this problem helps to clarify computational mechanisms underlying reservoir computing~\cite{Eoc-1990,Eoc-2004,Maass-2009} and brain dynamics modeled by recurrent neural networks~\cite{Vyas-2020,Ostojic-2024}.

Here, we provide concise and strong evidence about where and how the high dimensional neural dynamics are distributed, by designing a thermodynamic potential and further providing an analytic solution. This is possible only when a quasi-potential can be defined for the non-gradient dynamics~\cite{Qiu-2024}. The recent breakthrough paves a distinct route toward a precise understanding of the high-dimensional geometry of neural dynamics. Our analytic calculation shows an M-shaped region in the phase space where the dynamics are confined, while the underlying physics is a fixed-point anchored potential. The M shape will gradually shift to an arch shape in the deep chaotic region. The impact of this finding on nonlinear dynamics in both physics and neuroscience is discussed at the end of this paper.  

\section{Model}
The random recurrent neural network (RNN) is defined as a population of $N$ non-reciprocally interacting neurons, whose dynamics are captured by the following first-order differential equations~\cite{Chaos-1988}:
\begin{equation}\label{DynEq}
    \frac{dx_{i}}{dt}=-x_{i}+\sum_{j=1}^{N}J_{ij}r_{j},
\end{equation}
where $x_{i}\left(t\right),\ i=1,2,\cdots,N$ indicates the synaptic currents, $r_j\equiv\phi(x_j)$ where $\phi(\cdot)$ is the current-rate transfer function,  $J_{ij}\sim\mathcal{N}\left(0,g^{2}/N\right)$ are identically independently distributed coupling between neurons $(i\neq j)$.  $g$ is the gain parameter, $\phi\left(x\right)=\tanh x$ and $J_{ii}=0$ in this paper. $\phi=-1$ represents the bottom level of firing rate.

Increasing the gain strength leads to a chaos transition at $g_c=1$~\cite{Chaos-1988}, which inspired lots of follow-up works about stimulus-dependent suppression of chaos~\cite{PRE-2010}, memory capacity~\cite{PRE-2011}, topological complexity~\cite{PRL-2013}, attractor dimensionality~\cite{Clark-2023} and impacts of background noise~\cite{Helias-2018}. Distinct from previous works, our current work addresses a fundamental question about the geometry of the phase space where the dynamics are confined. A recent work conjectured that unstable fixed points and the dynamics are confined to separate shells~\cite{Helias-2022}, based on an analytic computation of the $\ell_2$ norm of neural activity. However, a single $\ell_2$ norm can not capture the internal structures of the confined region, and thus a direct mechanistic interpretation through an intuitive geometry is still lacking.

 We remark that to derive an entire picture of where the dynamics are confined, a distinct route must be established. First, we have to introduce the quasi-potential describing non-equilibrium steady states, and then define a geometry-oriented potential as a second key step, which we shall detail in the following.
 
 \section{Quasi-potential method}
Our previous work suggests that in the zero-speed limit (i.e., $d\bx/dt\to 0$), the RNN dynamics can be approximated by the following Langevin dynamics~\cite{Qiu-2024}
\begin{equation}
    \frac{dx_{i}}{dt}=-\frac{\partial E\left(\bx\right)}{\partial x_{i}}+\sqrt{2T}\epsilon_{i}\left(t\right),
\end{equation}
where $T$ is the temperature for the optimization of an energy function and finally tends to zero, and $\epsilon_i\left(t\right)$ is the Gaussian white noise. This energy function $E(\bx)$ is not the true Lyapunov function (may not exist) for the dynamics [Eq.~\eqref{DynEq}], but the quasi-potential (kinetic energy) specified below:
\begin{equation}
    E\left(\bx\right)=\frac{1}{2}\sum_{i}\left(-x_{i}+\sum_{j}J_{ij}\phi\left(x_{j}\right)\right)^{2}.
\end{equation}

In other words, minimizing the above quasi-potential via the Langevin dynamics amounts to searching for the fixed points of the original non-gradient dynamics.
Different temperatures control different levels of dynamics speed, e.g., $T\to 0$ achieving the ground state of $E$ (only fixed points are selected). The stationary limit of the Langevin dynamics captures the supporting bed of recurrent dynamics, i.e., the fixed points of Eq.~\eqref{DynEq} follow the canonical Boltzmann distribution
\begin{equation}\label{partF}
    p\left(\bx\right)=\frac{1}{Z} e^{-\beta E\left(\bx\right)},
\end{equation}
where the partition function $Z=\int d\bx e^{-\beta E(\bx)}$, and the inverse temperature $\beta=1/T\to\infty$. In this sense, we get rid of calculating the Kac-Rice formula~\cite{Kac,Rice}, which was used to count the number of roots for non-linear equations. This formula is hard to compute in most cases. Instead, we focus directly on the steady states of non-equilibrium dynamics with a large value of $\beta$, which is intuitively captured by the quasi-potential and the associated statistical mechanics of optimization. 

  \section{Thermodynamic potential of slow-point geometry}
  To define a geometry-oriented potential, we first slice the slow part of the phase space into different levels, and the bottom level corresponds to the ground state of the quasi-potential ($E(\bx)=0$). Then a typical reference point subject to $p(\bx)$, namely $\bx$, is picked up, and a constrained partition function is immediately defined below:
  \begin{equation}
    \Omega\left(d;\bx\right)=\int d\by\ e^{-\tilde{\beta}E\left(\by\right)}\delta\left(d^{2}-\frac{1}{N}\Vert\by-\bx\Vert^{2}\right),
\end{equation}
where the second inverse temperature $\tilde{\beta}$ tunes the different levels of the dynamics speed. In the limit $\tilde{\beta}\to\infty$, $\Omega\left(d;\bx\right)$ gives the volume of fixed points lying at the Euclidean distance $\sqrt{N}d$ from the reference $\bx$. In this sense, a distance-dependent entropy can be calculated as
$ s\left(d;\bx\right)=\frac{1}{N}\ln \Omega\left(d;\bx\right)$. In fact, the reference point $\bx$ follows the above Boltzmann distribution [Eq.~\eqref{partF}]. Therefore, we need to compute an average over this distribution to remove the dependence of the local entropy on $\bx$, as well as the coupling distribution, leading to the following concise expression:
\begin{equation}\label{LE}
    s\left(d\right)=-\tilde{\beta}\mathbb{E}_{\mathbf{J}}V,
\end{equation}
where
\begin{equation}
    V=-\frac{1}{N\tilde{\beta}}\left(\int d\bx\ e^{-\beta E\left(\bx\right)}\right)^{-1}\int d\bx\ e^{-\beta E\left(\bx\right)}\ln\Omega\left(d;\bx\right).
\end{equation}

 The potential $V$ is also known as the seminal Franz-Parisi potential, which was first introduced to characterize the glass transition in the $p$-spin spherical model~\cite{Franz-1995}, and was later elaborated to clarify the isolated-solution landscape in the context of the binary perceptron~\cite{PRE-2014}, and further led to the discovery of the rare dense solutions in neural networks~\cite{Baldassi-2015}. However, the geometry analysis in high dimensional non-equilibrium dynamics is extremely challenging, as the Lyapunov function is commonly believed to be unavailable. Thanks to the recently introduced quasi-potential method~\cite{Qiu-2024}, we are now able to turn this challenging problem into an amenable equilibrium setting, based on the above physics intuition.

\section{Results}
\subsection{Order parameters determine the geometry shape of dynamically slow points}
The average free-entropy $s(d)$ can be calculated by using the replica method in physics~\cite{Mezard-1987,Huang-2022}. In essence, two mathematical identities---
$Z^{-1}=\lim_{m\to 0}Z^{m-1}$ and $ \ln\Omega=\lim_{n\to 0}\partial_{n}\Omega^{n}$, are used to introduce two sets of replicas: $\bx^{a},\ a=1,2,\cdots,m$ and $\by^{\mu},\ \mu=1,2,\cdots,n$. Therefore, the free-entropy reads
\begin{equation}
    s=\frac{1}{N}\lim_{m,n\to 0}\partial_{n}\mathbb{E}_{\mathbf{J}}\mathcal{Z}^{m,n},
\end{equation}
where
\begin{widetext}
    \begin{equation}\label{Eq. Zmn}
        \mathcal{Z}^{m,n}=\int\left(\prod_{a=1}^{m}d{\bx}^{a}\right)\left(\prod_{\mu=1}^{n}d{\by}^{\mu}\right)\ \exp\left(-\beta \sum_{a}E\left({\bx}^{a}\right)-\tilde{\beta}\sum_{\mu}E\left({\by}^{\mu}\right)\right)\prod_{\mu}\delta\left(d^{2}-\frac{1}{N}\Vert{\by}^{\mu}-{\bx}^{1}\Vert^{2}\right),
    \end{equation}
\end{widetext}
 where we choose the reference fixed point to be $\bx^1$ without loss of generality. 
 
 After completing the average over two sources of randomness (reference point and coupling), we find that different neurons are decoupled, while different replicas are coupled.
We then have to introduce the following overlap parameters (order parameters in physics):
\begin{equation}
    \begin{aligned}
        &Q^{ab}=\frac{1}{N}\sum_{i}\phi\left({x}_{i}^{a}\right)\phi\left({x}_{i}^{b}\right),\ a\le b\\ 
        &\tilde{Q}^{\mu\nu}=\frac{1}{N}\sum_{i}\phi\left({y}_{i}^{\mu}\right)\phi\left({y}_{i}^{\nu}\right),\ \mu\le\nu\\ 
        &O^{a\mu}=\frac{1}{N}\sum_{i}\phi\left({x}_{i}^{a}\right)\phi\left({y}_{i}^{\mu}\right).
    \end{aligned}
\end{equation}
Intuitively, $Q$ and $\tilde{Q}$ describe the spaces of fixed-point reference ($\beta\to\infty$) and slow points ($\tilde{\beta}$ is kept finite), respectively. The appearance of $O$ is due to the distance constraint. 
A lengthy calculation is presented in the supplemental material (SM), together with the associated action function and saddle point equations (SDEs) optimizing the action. We only discuss the main physics in the main text.

 To get a physics result, we use the replica symmetry (RS) ansatz, i.e., the permutation of the replica index does not affect the physics. This is a first-level approximation, checked self-consistently by the stability of the SDEs. Explaining the distance constraint by a Fourier representation of the Dirac function leads to an extra conjugate quantity, namely $\hat{p}$, in the RS ansatz. Then, the slope of the free-entropy can be derived as $\frac{\partial s}{\partial d}=-2\hat{p}d$, where $\hat{p}$ is determined by solving the SDEs (see SM). This equality is a fundamental equation that determines the shape of the phase space confining the slow points of the dynamics.
 \begin{figure}
    \centering
        \includegraphics[scale=0.5]{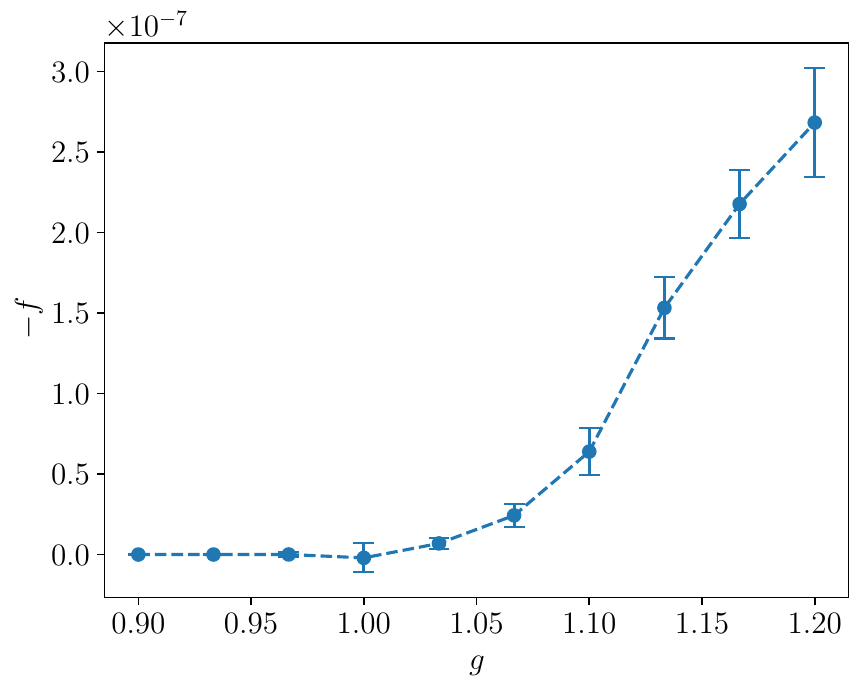}
    \caption{The free energy of the reference has a transition at $g_c=1$.}\label{free}
\end{figure}

 \begin{figure}
    \centering
    \subfigure[]{
        \includegraphics[scale=0.5]{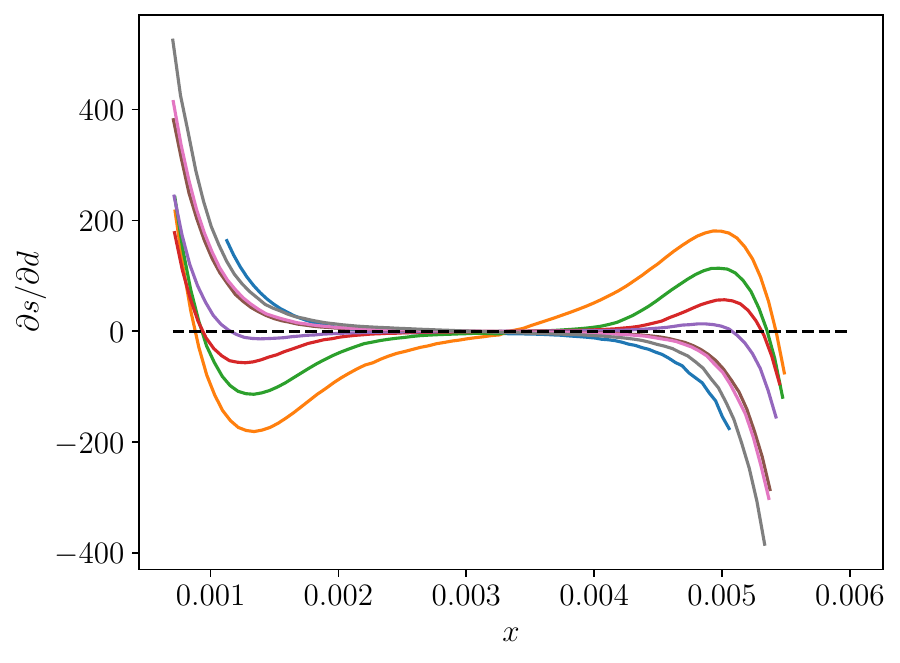}
   }
   \subfigure[]{
        \includegraphics[scale=0.5]{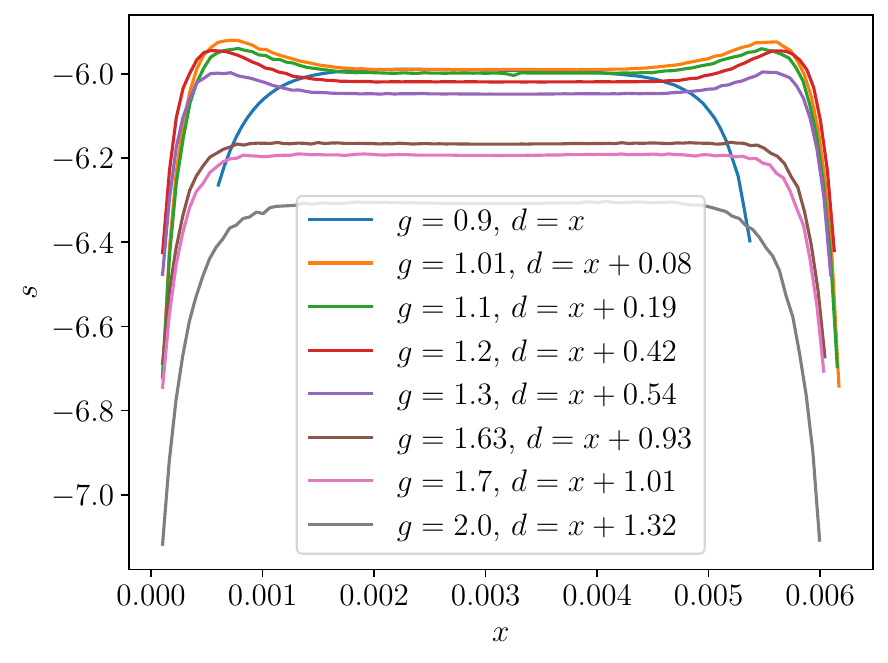}
       }
    \caption{The geometry of the phase pace confining high-dimensional recurrent dynamics. Local entropy (b) and its derivative (a) as a function of the distance for different values of $g$. If $g<1$, the local entropy profile is a spike near the origin. If $g>1$, the profile shows an M-shaped structure and then arched structure, and the corresponding position becomes further away from the reference with increasing $g$.}\label{LEP}
\end{figure}

 \subsection{M-shaped and arched regions confining slow points of dynamics}
 We first consider the free energy of all unstable fixed points in Fig.~\ref{free} [see Eq.~\eqref{partF}], revealing that the free energy vanishes before the onset of chaos, and starts to increase after the chaos.
 This demonstrates that the number of fixed points explodes in the chaotic regime, consistent with previous works~\cite{PRL-2013,Helias-2022}. Next we show what were missed in previous
 theoretical studies.
 
By solving the SDEs (see SM), one can clarify the shape of the phase space confining the dynamics at different levels of slowness. The
local entropy profile and its derivative are shown in Fig.~\ref{LEP}. If $g$ is less than the critical value $g_c=1$, where the trivial null activity is a global fixed point of the dynamics~\cite{Chaos-1988},  the local entropy profile is a spike at $d\sim 0$, which is consistent with the known picture that in the ordered (non-chaotic) phase the dominant solution for the non-gradient recurrent dynamics is the all-zero activities. Because we take $\tilde{\beta}=10^{6}$, the spike is very close to the trivial null point. This spike shape can also be deduced from the derivative profile of the local entropy. The slope is a monotonically decreasing function of the separated distance from the trivial fixed point ($\beta\to\infty$ and $g<1$). 

We next look at the geometric interpretation of the chaos transition. Once the gain parameter crosses the critical value from below, there appears a qualitative change of the slope profile. The slope decreases abruptly from a positive infinite value and then rapidly increases until a plateau at the zero slope, after that the slope increases sharply again before finally dropping down to a negative infinite value. This distinct behavior is shared only in the shallow chaotic region, in stark contrast to the behavior in the non-chaotic region. This slope behavior determines the entire shape of the region confining dynamics. By our theoretical analysis, the internal shape of this phase space can thus be revealed---sharp boundaries exist, and in the middle, the slow-point density drops slightly, displaying a flat region. We thus call this an M-shaped confinement. With increasing the synaptic gain strength, another arch shape of the confinement gradually emerges at $g\simeq1.63$. The recent work~\cite{Helias-2022} is qualitatively consistent with our theory, yet missed the important internal structure of the phase space. Surprisingly, this recent work is based on the large deviation principle and random matrix theory, while our calculation is based on the quasi-potential and the geometry-oriented measure. It is thus very interesting for mathematicians to prove the coherence of such different theoretical lenses. 

\begin{figure}
    \centering
        \includegraphics[scale=0.5]{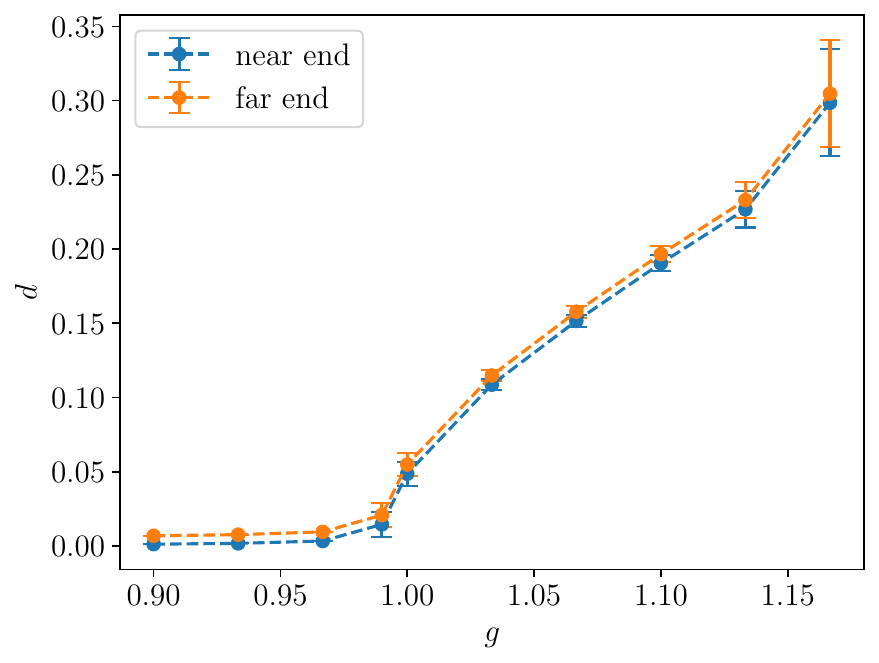}
    \caption{The confinement is repelled from the reference when increasing the gain parameter.}\label{width}
\end{figure}

\begin{figure}
    \centering
        \includegraphics[scale=0.5]{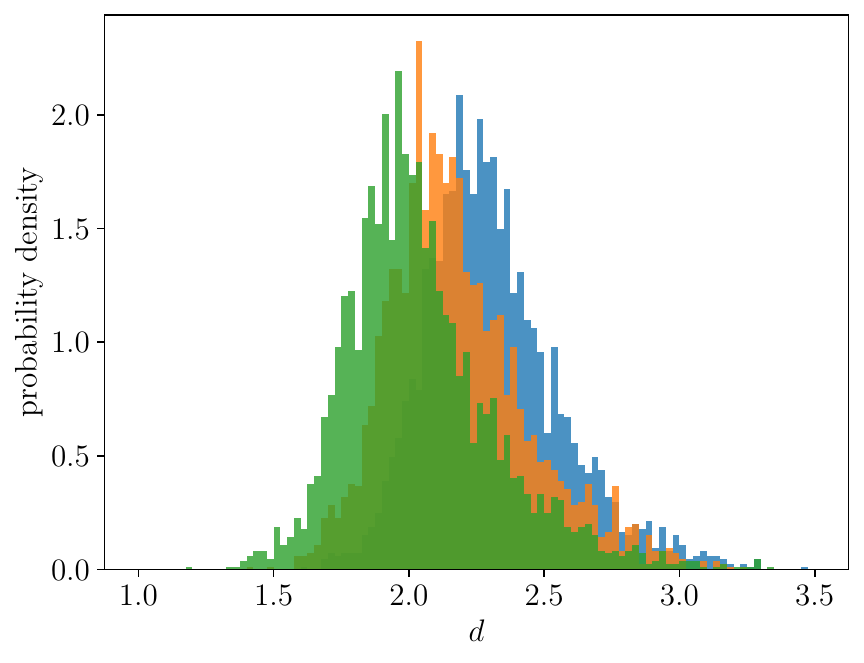}
    \caption{Distribution of low-speed solutions ($\|\dot{\bx}\|<10^{-3}$) around a fixed point ($\|\dot{\bx}\|<10^{-6}$) for network dynamics with $N=100$ and $g=3.0$. Each histogram represents the distribution around a specific reference. }\label{dist}
\end{figure}

We finally ask how the width and position of the confinement change with the gain parameter (Fig.~\ref{width}). Anchored at the fixed point (zero speed), the slower dynamics are narrowly distributed and separated from the anchor, but at the same level of slowness, the confinement gets further from the anchor with increasing values of $g$, and meanwhile, the corresponding space confining the dynamics becomes first M-shape and then arch shape yet with nearly constant width, which can not be intuitively implied from the fact that the number of unstable fixed points grows rapidly in the deep chaotic regime~\cite{Helias-2022}. Therefore, our theory not only reveals where the dynamics at different levels of slowness are confined but also reveals the entire internal structure of the phase space confining these dynamics. The theory is also numerically supported by the fixed-point finder algorithm working on finite-sized recurrent neural networks, e.g., the number of fixed points with a higher speed separated with a specified distance from the reference of a lower speed (see Fig.~\ref{dist} and technical details in SM).

\section{Concluding remarks}
Recurrent dynamics is ubiquitous in brains and neuro-inspired computing. A fundamental question is then where and how these recurrent dynamics are confined in the phase space, which is also a long-lasting unsolved question in high-dimensional dynamics. In this paper, we provide a semi-rigorous answer to this question and prove through analytic theory that an M-shaped and then arch-shaped region arises in the phase space when the chaos comes into play. This region bears sharp boundaries and a flat internal structure, characterized by a single physics potential, namely local entropy. The underlying mathematics is thus revealed for the high dimensional chaos, particularly in the context of random recurrent neural networks we consider here. 

This theoretical result marks an important advance over previous works showing the complexity and dimensionality of chaotic attractors~\cite{PRL-2013,Clark-2023,Helias-2022}. Based on this step, many fundamental questions can be asked and potentially addressed, e.g., how anatomical and physiological constraints reshape the dynamics geometry, what is the space of dynamics faced with external stimuli for cognitive tasks, and how the phase space is re-organized by learning (considering various kinds of synaptic plasticity in the brain). The theory can also be potentially extended to other types of dynamics, e.g., those studied in ecosystems~\cite{Ros-2023}. On the mathematical physics aspect, our formula also opens a promising route toward establishing a rigorous connection among different analytic concepts for high dimensional dynamics, such as dynamical complexity, topological complexity, large deviation principle, and random matrix theory.

\section{Acknowledgments}
This research was supported by the National Natural Science Foundation of China for
Grant number 12122515, and Guangdong Provincial Key Laboratory of Magnetoelectric Physics and Devices (No. 2022B1212010008), and Guangdong Basic and Applied Basic Research Foundation (Grant No. 2023B1515040023).

\onecolumngrid
\appendix

\section{Theoretical details of replica calculation}
In this appendix, we provide details of the replica method to compute the free-entropy of the dynamics,  a seminal method in disordered system theory~\cite{Mezard-1987}. The disorder in our model comes from the choice of the reference point and the network couplings. The Franz-Parisi potential must be averaged over these quenched disorders. The replica method is based on the following two mathematical identities: $Z^{-1}=\lim_{m\to 0}Z^{m-1}$ and $ \ln\Omega=\lim_{n\to 0}\partial_{n}\Omega^{n}$. Therefore, two sets of replica indexes are introduced as $\bx^{a},\ a=1,2,\cdots,m$ and $\by^{\mu},\ \mu=1,2,\cdots,n$.

We first write $ \mathbb{E}_{\mathbf J} \mathcal{Z}^{m,n}$ in the main text in an explicit form:
\begin{equation}
\begin{aligned}
    &\mathbb{E}_{\mathbf{J}} \mathcal{Z}^{m,n}=\int\left(\prod_{ia} D\hat{{x}}_{i}^{a}d{x}_{i}^{a}\right)\left(\prod_{i\mu} D\hat{{y}}_{i}^{\mu}d{y}_{i}^{\mu}\right)\exp\left(-\sqrt{\beta}\sum_{ia}i\hat{x}_{i}^{a}x_{i}^{a}-\sqrt{\tilde{\beta}}\sum_{i\mu}i\hat{y}_{i}^{\mu}y_{i}^{\mu}\right)\\
    & \left[\prod_{\mu}\delta\left(d^{2}-\frac{1}{N}\sum_{i}\left({y}_{i}^{\mu}-{x}_{i}^{1}\right)^{2}\right)\right]\mathbb{E}_{\mathbf{J}} \exp\left(\sqrt{\beta }\sum_{ia}i\hat{x}_{i}^{a}\sum_{j}{J}_{ij}\phi\left({x}_{j}^{a}\right)+\sqrt{\tilde{\beta}}\sum_{i\mu}i\hat{y}_{i}^{\mu}\sum_{j}{J}_{ij}\phi\left({y}_{j}^{\mu}\right)\right),
\end{aligned}
\end{equation}
where the sum over $i$ refers only to the subscript (neuron index), and the following Hubbard-Stratonovich transformation has been performed to linearize the quadratic terms in the exponential function.
\begin{equation}
    e^{-ax^{2}/2}=\int D\hat{x}\ e^{i\sqrt{a}\hat{x}x},
\end{equation}
where $a>0$, $D\hat{x}=\frac{d\hat{x}}{\sqrt{2\pi}}e^{-\hat{x}^{2}/2}$ is a Gaussian measure.

The disorder average over the coupling can be easily worked out as follows,
\begin{equation}\label{DA}
    \begin{aligned}
        &\mathbb{E}_{\mathbf{J}} \exp\left(\sqrt{\beta }\sum_{ia}i\hat{x}_{i}^{a}\sum_{j}{J}_{ij}\phi\left({x}_{j}^{a}\right)+\sqrt{\tilde{\beta}}\sum_{i\mu}i\hat{y}_{i}^{\mu}\sum_{j}{J}_{ij}\phi\left({y}_{j}^{\mu}\right)\right)\\
        \approx&\int\left(\prod_{ij} D{z}_{ij}\right)\ \exp\left[\sqrt{\frac{g^{2}}{N}}\sum_{ij}z_{ij}\left(\sqrt{\beta}\sum_{a}i\hat{x}_{i}^{a}\phi\left(x_{j}^{a}\right)+\sqrt{\tilde{\beta}}\sum_{\mu}i\hat{y}_{i}^{\mu}\phi\left(y_{j}^{\mu}\right)\right)\right]\\
        =&\exp\Bigg\{\frac{1}{2}g^{2}\Bigg[\beta\sum_{ab}\left(\sum_{i}i\hat{{x}}_{i}^{a}i\hat{{x}}_{i}^{b}\right)\left(\frac{1}{N}\sum_{i}\phi\left({x}_{i}^{a}\right)\phi\left({x}_{i}^{b}\right)\right)+\tilde{\beta}\sum_{\mu\nu}\left(\sum_{i}i\hat{{y}}_{i}^{\mu}i\hat{{y}}_{i}^{\nu}\right)\left(\frac{1}{N}\sum_{i}\phi\left({y}_{i}^{\mu}\right)\phi\left({y}_{i}^{\nu}\right)\right)\\
        &\ +2\sqrt{\beta\tilde{\beta}}\sum_{a\mu}\left(\sum_{i}i\hat{{x}}_{i}^{a}i\hat{{y}}_{i}^{\mu}\right)\left(\frac{1}{N}\sum_{i}\phi\left({x}_{i}^{a}\right)\phi\left({y}_{i}^{\mu}\right)\right)\Bigg]\Bigg\},
    \end{aligned}
\end{equation}
where the diagonal elements $J_{ii}$ are inserted, but their effects are negligible in the thermodynamic limit~\cite{Qiu-2024}. 

A careful inspection of Eq.~\eqref{DA} reveals that one can replace three normalized summations over neuron index by the order parameters defined in the main text, and therefore three integrals of Dirac delta function are required to be inserted. In addition, the Dirac delta functions to enforce these order parameters can be expressed by their Fourier representations, thereby introducing the conjugate order parameters (see hatted quantities below). We thus arrive at the following concise expression:
\begin{equation}
\begin{aligned}
    \mathbb{E}_{J} \mathcal{Z}^{m,n}&=\int\left(\prod_{a\le b}\frac{d\hat{Q}^{ab}}{2\pi i/N}dQ^{ab}\right)\left(\prod_{\mu\le\nu}\frac{d\hat{\tilde{Q}}^{\mu\nu}}{2\pi i/N}d\tilde{Q}^{\mu\nu}\right)\left(\prod_{a\mu}\frac{d\hat{O}^{a\mu}}{2\pi i/N}dO^{a\mu}\right)\\
    &\left(\prod_{\mu}\frac{d\hat{p}^{\mu}}{2\pi i/N}\right)\exp\left[N\left(G_{0}+G\right)\right],
    \end{aligned}
\end{equation}
where
\begin{equation}
    G_{0}=-\sum_{a\le b}\hat{Q}^{ab}Q^{ab}-\sum_{\mu\le\nu}\hat{\tilde{Q}}^{\mu\nu}\tilde{Q}^{\mu\nu}-\sum_{a\mu}\hat{O}^{a\mu}O^{a\mu}-d^{2}\sum_{\mu}\hat{p}^{\mu},
\end{equation}
and
\begin{equation}
\begin{aligned}
    G=&\ln\int\left(\prod_{a} D\hat{x}^{a}dx^{a}\right)\left(\prod_{\mu} D\hat{y}^{\mu}dy^{\mu}\right)\exp\Bigg[-\sqrt{\beta }\sum_{a}i\hat{x}^{a}x^{a}-\sqrt{\tilde{\beta}}\sum_{\mu}i\hat{y}^{\mu}y^{\mu}\notag\\
    &\ +\frac{1}{2}g^{2}\left(\beta \sum_{ab}i\hat{x}^{a}i\hat{x}^{b}Q^{ab}+\tilde{\beta}\sum_{\mu\nu}i\hat{y}^{\mu}i\hat{y}^{\nu}\tilde{Q}^{\mu\nu}+2\sqrt{\beta\tilde{\beta} }\sum_{a\mu}i\hat{x}^{a}i\hat{y}^{\mu}O^{a\mu}\right)\notag\\
    &\ +\sum_{a\le b}\hat{Q}^{ab}\phi\left(x^{a}\right)\phi\left(x^{b}\right)+\sum_{\mu\le\nu}\hat{\tilde{Q}}^{\mu\nu}\phi\left(y^{\mu}\right)\phi\left(y^{\nu}\right)+\sum_{a\mu}\hat{O}^{a\mu}\phi\left(x^{a}\right)\phi\left(y^{\mu}\right)+\sum_{\mu}\hat{p}^{\mu}\left(y^{\mu}-x^{1}\right)^{2}\Bigg].
\end{aligned}
\end{equation}
Note that $\hat{p}^\mu$ is introduced due to the Fourier representation of the Dirac delta function enforcing the distance constraint. The sum of $G_0+G$ is called the action in physics. The geometry information of the dynamics can be read out from the stationary condition of this action in the thermodynamic limit.

\subsection{Replica symmetric ansatz}
The replica symmetric ansatz can be mathematically expressed as follows,
\begin{equation}
    \begin{aligned}
        &Q^{ab}=Q\left(1-\delta_{ab}\right)+q\delta_{ab},\\
        &\tilde{Q}^{\mu\nu}=\tilde{Q}\left(1-\delta_{\mu\nu}\right)+\tilde{q}\delta_{\mu\nu},\\
        &O^{a\mu}=O\left(1-\delta_{a1}\right)+o\delta_{a1},\\
        &\hat{p}^{\mu}=\hat{p}.
    \end{aligned}
\end{equation} 
The free part of the action can be simplified as follows,
\begin{equation}
    G_{0}=-\frac{1}{2}m\left(m-1\right)\hat{Q}Q-m\hat{q}q-\frac{1}{2}n\left(n-1\right)\hat{\tilde{Q}}\tilde{Q}-n\hat{\tilde{q}}\tilde{q}-n\left(m-1\right)\hat{O}O-n\hat{o}o-n\hat{p}d^{2}.
\end{equation}
The interaction part is a bit complicated, but tractable by repeated applications of the Hubbard-Stratonovich transformation. The details are given below.
\begin{equation}\label{Gdet}
    \begin{aligned}
        G=&\ln\int\left(\prod_{a} D\hat{x}^{a}dx^{a}\right)\left(\prod_{\mu} D\hat{y}^{\mu}dy^{\mu}\right)\exp\Bigg[-\sqrt{\beta }\sum_{a}i\hat{x}^{a}x^{a}-\sqrt{\tilde{\beta}}\sum_{\mu}i\hat{y}^{\mu}y^{\mu}\\
        & +\frac{1}{2}g^{2}\Bigg\{\beta\left(Q\left(\sum_{a}i\hat{x}^{a}\right)^{2}+\left(q-Q\right)\sum_{a}\left(i\hat{x}^{a}\right)^{2}\right)+\tilde{\beta}\left(\tilde{Q}\left(\sum_{\mu}i\hat{y}^{\mu}\right)^{2}+\left(\tilde{q}-\tilde{Q}\right)\sum_{\mu}\left(i\hat{y}^{\mu}\right)^{2}\right)\\
        &+2\sqrt{\beta\tilde{\beta}}\Bigg[\frac{1}{2}O\left(\left(\sum_{a}i\hat{x}^{a}+\sum_{\mu}i\hat{y}^{\mu}\right)^{2}-\left(\sum_{a}i\hat{x}^{a}\right)^{2}-\left(\sum_{\mu}i\hat{y}^{\mu}\right)^{2}\right)+\left(o-O\right)i\hat{x}^{1}\sum_{\mu}i\hat{y}^{\mu}\Bigg]\Bigg\}\\
        & +\frac{1}{2}\hat{Q}\left(\sum_{a}\phi\left(x^{a}\right)\right)^{2}+\left(\hat{q}-\frac{1}{2}\hat{Q}\right)\sum_{a}\phi^{2}\left(x^{a}\right)+\frac{1}{2}\hat{\tilde{Q}}\left(\sum_{\mu}\phi\left(y^{\mu}\right)\right)^{2}+\left(\hat{\tilde{q}}-\frac{1}{2}\hat{\tilde{Q}}\right)\sum_{\mu}\phi^{2}\left(y^{\mu}\right)\\
        & +\frac{1}{2}\hat{O}\left(\left(\sum_{a}\phi\left(x^{a}\right)+\sum_{\mu}\phi\left(y^{\mu}\right)\right)^{2}-\left(\sum_{a}\phi\left(x^{a}\right)\right)^{2}-\left(\sum_{\mu}\phi\left(y^{\mu}\right)\right)^{2}\right)\\
        & +\left(\hat{o}-\hat{O}\right)\phi\left(x^{1}\right)\sum_{\mu}\phi\left(y^{\mu}\right)+\hat{p}\sum_{\mu}\left(y^{\mu}-x^{1}\right)^{2}\Bigg]\\
        =&\ln\int\left(\prod_{a}dx^{a}\right)\ \exp\left(\sum_{a}\left(\hat{q}-\frac{1}{2}\hat{Q}\right)\phi^{2}\left(x^{a}\right)\right)\\
        & \int\left(\prod_{a} D\hat{x}^{a}\right)\ \exp\left(\sum_{a}\left(\frac{1}{2}g^{2}\beta\left(q-Q\right)\left(i\hat{x}^{a}\right)^{2}-\sqrt{\beta}x^{a}i\hat{x}^{a}\right)\right)\\
        &\times \int\left(\prod_{\mu}dy^{\mu}\right)\exp\Bigg[\sum_{\mu}\Bigg(\left(\hat{\tilde{q}}-\frac{1}{2}\hat{\tilde{Q}}\right)\phi^{2}\left(y^{\mu}\right)+\left(\hat{o}-\hat{O}\right)\phi\left(x^{1}\right)\phi\left(y^{\mu}\right)+\hat{p}\left(y^{\mu}-x^{1}\right)^{2}\Bigg)\Bigg]\\
        &\times \int\left(\prod_{\mu} D\hat{y}^{\mu}\right)\ \exp\left[\sum_{\mu}\Bigg(\frac{1}{2}g^{2}\tilde{\beta}\left(\tilde{q}-\tilde{Q}\right)\left(i\hat{y}^{\mu}\right)^{2}+\left(-\sqrt{\tilde{\beta}}y^{\mu}+g^{2}\sqrt{\beta\tilde{\beta}}\left(o-O\right)i\hat{x}^{1}\right)i\hat{y}^{\mu}\Bigg)\right]\\
        &\times \int D{z}\ \exp\left(u_{1}\sum_{a}\phi\left(x^{a}\right)+u_{2}\sum_{a}i\hat{x}^{a}+u_{3}\sum_{\mu}\phi\left(y^{\mu}\right)+u_{4}\sum_{\mu}i\hat{y}^{\mu}\right)\\
        =&\ln\int D{z}\ \left(\int dx\ e^{H}\right)^{m-1}\int dx\ \exp\left[\left(\hat{q}-\frac{1}{2}\hat{Q}\right)\phi^{2}\left(x\right)+u_{1}\phi\left(x\right)\right]\\
        &\times \int D\hat{x}\ \exp\left(\frac{1}{2}g^{2}\beta\left(q-Q\right)\left(i\hat{x}\right)^{2}+\left(-\sqrt{\beta}x+u_{2}\right)i\hat{x}\right)\left(\int dy\ e^{\tilde{H}}\right)^{n},
    \end{aligned}
\end{equation}
where $Dz\equiv Dz_1Dz_2Dz_3Dz_4$ and
\begin{subequations}\label{paru}
\begin{align}
    u_{1}&=z_{1}\sqrt{\hat{Q}},\ u_{2}=z_{2}g\sqrt{\beta Q},\\
    u_{3}&=z_{1}\frac{\hat{O}}{\sqrt{\hat{Q}}}+z_{3}\sqrt{\hat{\tilde{Q}}-\frac{\hat{O}^{2}}{\hat{Q}}},\ u_{4}=g\sqrt{\tilde{\beta}}\left(z_{2}\frac{O}{\sqrt{Q}}+z_{4}\sqrt{\tilde{Q}-\frac{O^{2}}{Q}}\right).
    \end{align}
\end{subequations}
Note that $\tilde{H}$ contains $u_3$ and $u_4$ (see below).

During derivation of the interaction part, we also derive two effective Hamiltonians. The outer Hamiltonian is specified below.
\begin{equation}
    \begin{aligned}
        H=&\ln\left\{\exp\left[\left(\hat{q}-\frac{1}{2}\hat{Q}\right)\phi^{2}\left(x\right)+u_{1}\phi\left(x\right)\right]\int D\hat{x}\ \exp\left(\frac{1}{2}g^{2}\beta\left(q-Q\right)\left(i\hat{x}\right)^{2}+\left(-\sqrt{\beta}x+u_{2}\right)i\hat{x}\right)\right\}\\
        =&\ln\Bigg(\exp\left[\left(\hat{q}-\frac{1}{2}\hat{Q}\right)\phi^{2}\left(x\right)+u_{1}\phi\left(x\right)\right]\frac{1}{\sigma}\exp\left(-\frac{1}{2\sigma^{2}}\left(-\sqrt{\beta}x+u_{2}\right)^{2}\right)\Bigg)\\
        =&\left(\hat{q}-\frac{1}{2}\hat{Q}\right)\phi^{2}\left(x\right)+u_{1}\phi\left(x\right)-\frac{1}{2\sigma^{2}}\left(-\sqrt{\beta}x+u_{2}\right)^{2}-\ln\sigma,
    \end{aligned}
\end{equation}
where $\sigma=\sqrt{1+g^{2}\beta\left(q-Q\right)}$. The inner Hamiltonian is specified as follows.
\begin{equation}
\begin{aligned}
    \tilde{H}=&\left(\hat{\tilde{q}}-\frac{1}{2}\hat{\tilde{Q}}\right)\phi^{2}\left(y\right)+u_{3}\phi\left(y\right)+\left(\hat{o}-\hat{O}\right)\phi\left(x\right)\phi\left(y\right)+\hat{p}\left(y-x\right)^{2}\notag\\
    &\ -\frac{1}{2\tilde{\sigma}^{2}}\left(-\sqrt{\tilde{\beta}}y+u_{4}+g^{2}\sqrt{\beta\tilde{\beta} }\left(o-O\right)i\hat{x}\right)^{2}-\ln\tilde{\sigma},
\end{aligned}
\end{equation}
where $\tilde{\sigma}=\sqrt{1+g^{2}\tilde{\beta}\left(\tilde{q}-\tilde{Q}\right)}$. 

To derive the integral $\int Dz$ in Eq.~\eqref{Gdet}, we have to re-express the following function into its integral representation:
\begin{equation}
    \begin{aligned}
        &\exp\Bigg[\frac{1}{2}\left(\hat{Q}-\hat{O}\right)\left(\sum_{a}\phi\left(x^{a}\right)\right)^{2}+\frac{1}{2}g^{2}\left(\beta Q-\sqrt{\beta\tilde{\beta}}O\right)\left(\sum_{a}i\hat{x}^{a}\right)^{2}\notag\\
        &\ +\frac{1}{2}\left(\hat{\tilde{Q}}-\hat{O}\right)\left(\sum_{\mu}\phi\left(y^{\mu}\right)\right)^{2}+\frac{1}{2}g^{2}\left(\tilde{\beta}\tilde{Q}-\sqrt{\beta\tilde{\beta}}O\right)\left(\sum_{\mu}i\hat{y}^{\mu}\right)^{2}\notag\\
        &\ +\frac{1}{2}\hat{O}\left(\sum_{a}\phi\left(x^{a}\right)+\sum_{\mu}\phi\left(y^{\mu}\right)\right)^{2}+\frac{1}{2}g^{2}\sqrt{\beta\tilde{\beta}}O\left(\sum_{a}i\hat{x}^{a}+\sum_{\mu}i\hat{y}^{\mu}\right)^{2}\Bigg]\\
        =&\int DtD\tau \exp\Bigg[t_{1}\sqrt{\hat{Q}-\hat{O}}\sum_{a}\phi\left(x^{a}\right)+t_{2}g\sqrt{\beta Q-\sqrt{\beta\tilde{\beta}}O}\sum_{a}i\hat{x}^{a}\notag\\
        &\ +t_{3}\sqrt{\hat{\tilde{Q}}-\hat{O}}\sum_{\mu}\phi\left(y^{\mu}\right)+t_{4}g\sqrt{\tilde{\beta}\tilde{Q}-\sqrt{\beta\tilde{\beta}}O}\sum_{\mu}i\hat{y}^{\mu}\notag\\
        &\ +\tau_{1}\sqrt{\hat{O}}\left(\sum_{a}\phi\left(x^{a}\right)+\sum_{\mu}\phi\left(y^{\mu}\right)\right)+\tau_{2}g\sqrt{\sqrt{\beta\tilde{\beta}}O}\left(\sum_{a}i\hat{x}^{a}+\sum_{\mu}i\hat{y}^{\mu}\right)\Bigg]\\
        =&\int Dz\exp\left(u_{1}\sum_{a}\phi\left(x^{a}\right)+u_{2}\sum_{a}i\hat{x}^{a}+u_{3}\sum_{\mu}\phi\left(y^{\mu}\right)+u_{4}\sum_{\mu}i\hat{y}^{\mu}\right),
    \end{aligned}
\end{equation}
where $Dt\equiv Dt_1Dt_2Dt_3Dt_4$, $D\tau\equiv D\tau_1D\tau_2D\tau_3D\tau_4$, and
\begin{gather}
    u_{1}=t_{1}\sqrt{\hat{Q}-\hat{O}}+\tau_{1}\sqrt{\hat{O}},\ u_{2}=t_{2}g\sqrt{\beta Q-\sqrt{\beta\tilde{\beta}}O}+\tau_{2}g\sqrt{\sqrt{\beta\tilde{\beta}}O}\notag\\ 
    u_{3}=t_{3}\sqrt{\hat{\tilde{Q}}-\hat{O}}+\tau_{1}\sqrt{\hat{O}},\ u_{4}=t_{4}g\sqrt{\tilde{\beta}\tilde{Q}-\sqrt{\beta\tilde{\beta}}O}+\tau_{2}g\sqrt{\sqrt{\beta\tilde{\beta}}O},
\end{gather}
which can be reparameterized by using $Dz$ that keeps the same statistics, leading to Eq.~\eqref{paru}.

\subsection{Zero-temperature limit}
Because our reference point is chosen to be an exact fixed point, $\beta$ should tend to an infinite value. In this limit,
the order parameters have the following scaling behavior with $\beta$:
\begin{equation}
    \begin{aligned}
        &Q\to q-\frac{1}{\beta}\delta Q,\\
        &O\to o-\frac{1}{\beta}\delta O.
    \end{aligned}
\end{equation}
 The conjugate order parameters behave in a corresponding fashion:
\begin{equation}
    \hat{q}\to \beta\delta\hat{q}+\frac{1}{2}\beta^{2}\hat{Q},\ \hat{Q}\to \beta^{2}\hat{Q},
\end{equation}
and 
\begin{equation}
    \hat{o}\to\delta\hat{o}+\beta\hat{O},\ \hat{O}\to\beta\hat{O}.
\end{equation}

In the large $\beta$ limit, we have
\begin{equation}
\begin{split}
    G_{0}&=\frac{1}{2}m\left(m-1\right)\beta\hat{Q}\delta Q-m\beta q\delta\hat{q}-\frac{1}{2}n\left(n-1\right)\hat{\tilde{Q}}\tilde{Q}-n\hat{\tilde{q}}\tilde{q}+n\left(m-1\right)\hat{O}\delta O\\
    &-n\delta\hat{o}o-n\hat{p}d^{2}-\frac{1}{2}m^{2}\beta^{2}\hat{Q}q-nm\beta\hat{O}o.
    \end{split}
\end{equation}
We also have the reduced outer Hamiltonian:
\begin{equation}
    H=\beta\delta\hat{q}\phi^{2}\left(x\right)+\beta z_{1}\sqrt{\hat{Q}}\phi\left(x\right)-\frac{\beta}{2\sigma^{2}}\left(-x+z_{2}g\sqrt{q}\right)^{2}-\ln\sigma,
\end{equation}
where $\sigma=\sqrt{1+g^{2}\delta Q}$. The inner Hamiltonian reads
\begin{equation}
    \begin{aligned}
        \tilde{H}=&\left(\hat{\tilde{q}}-\frac{1}{2}\hat{\tilde{Q}}\right)\phi^{2}\left(y\right)+\left(z_{1}\frac{\hat{O}}{\sqrt{\hat{Q}}}+z_{3}\sqrt{\hat{\tilde{Q}}-\frac{\hat{O}^{2}}{\hat{Q}}}\right)\phi\left(y\right)+\delta\hat{o}\phi\left(x\right)\phi\left(y\right)+\hat{p}\left(y-x\right)^{2}\\
        &\ -\frac{1}{2\tilde{\sigma}^{2}}\left[-\sqrt{\tilde{\beta}}y+g\sqrt{\tilde{\beta}}\left(z_{2}\frac{o}{\sqrt{q}}+z_{4}\sqrt{\tilde{Q}-\frac{o^{2}}{q}}\right)\right]^{2}-\ln\tilde{\sigma},
    \end{aligned}
\end{equation}
where the original $i\hat{x}$ term is now absent in the limit $\beta\to\infty$, and thus
\begin{equation}
    G=\ln\int Dz\ \left(\int dx\ e^{H}\right)^{m-1}\int dx\ e^{H}\left(\int dy\ e^{\tilde{H}}\right)^{n}.
\end{equation}

Keeping up to the first order of $m$, the extremization of $G_{0}+G$ with respect to $\mathcal{O},\hat{\mathcal{O}}=\{q,\delta Q,\delta\hat{q},\hat{Q}\}$ in the zero-replica limit ($m,n\to 0$) gives the self-consistent saddle-point equations (SDEs). More precisely,
\begin{equation}
    \partial_{\mathcal{O},\hat{\mathcal{O}}}\left(\mathcal{G}_{0}+\mathcal{G}\right)=0,
\end{equation}
where
\begin{equation}
    \mathcal{G}_{0}=\lim_{m,n\to 0}\partial_{m}G_{0}=-\frac{1}{2}\beta\hat{Q}\delta Q-\beta q\delta\hat{q},
\end{equation}
and
\begin{equation}
    \begin{aligned}
        \mathcal{G}=&\lim_{m,n\to 0}\partial_{m}G\\
        =&\int Dz\ \ln\int dx\ e^{H}\\
        =&-\frac{\beta g^{2}}{2\sigma^{2}}q-\beta\int Dz\ \mathcal{H}\left(x^{*}\right),
    \end{aligned}
\end{equation}
where
\begin{equation}\label{Hfun}
    \mathcal{H}\left(x\right)=\left(\frac{1}{2\sigma^{2}}\right)x^{2}-\delta\hat{q}\phi^{2}\left(x\right)-z_{1}\sqrt{\hat{Q}}\phi\left(x\right)-\frac{z_{2}g\sqrt{q}}{\sigma^{2}}x,
\end{equation}
and $x^{*}={\rm argmin}\ \mathcal{H}\left(x\right)$.

After a cumbersome calculation, we get the following SDEs:
\begin{subequations}\label{Eq. SDEs q}
    \begin{align}
        &q=[\phi^{2}\left(x^{*}\right)],\\
        &\delta Q=\frac{1}{\sqrt{\hat{Q}}}[z_{1}\phi\left(x^{*}\right)],\\
        &\delta\hat{q}=-\frac{g^{2}}{2\sigma^{2}}+\frac{g}{2\sigma^{2}\sqrt{q}}[z_{2}x^{*}],\\
        &\hat{Q}=\frac{g^{4}q}{\sigma^{4}}+\frac{g^{2}}{\sigma^{4}}[{x^{*}}^{2}]-\frac{2g^{3}\sqrt{q}}{\sigma^{4}}[z_{2}x^{*}],
    \end{align}
\end{subequations}
where $[\cdot]=\int Dz\ \cdot$. This is exactly the SDEs presented in the previous work~\cite{Qiu-2024}, as expected from the fact that the overlap matrix $Q^{ab}$ captures the statistics in the space of reference fixed points.

Similarly, keeping up to the first order of $n$ for the operation $\partial_n$, and computing the derivatives with respect to $\tilde{\mathcal{O}},\hat{\tilde{\mathcal{O}}}=\{\tilde{q},\tilde{Q},o,\hat{\tilde{q}},\hat{\tilde{Q}},\delta\hat{o}\}$ produces another set of SDEs, which also relies on the order parameters of the reference space. More precisely,
\begin{equation}
    \partial_{\tilde{\mathcal{O}},\hat{\tilde{\mathcal{O}}}}\left(\tilde{\mathcal{G}}_{0}+\tilde{\mathcal{G}}\right)=0,
\end{equation}
where
\begin{subequations}
\begin{align}
    \tilde{\mathcal{G}}_{0}&=\frac{1}{2}\hat{\tilde{Q}}\tilde{Q}-\hat{\tilde{q}}\tilde{q}-\hat{O}\delta O-\delta\hat{o}o-\hat{p}d^{2},\\
    \tilde{\mathcal{G}}&=\frac{1}{2}\ln \frac{\tilde{k}}{g\tilde{\beta}}-\frac{1}{2}g\tilde{k}\tilde{Q}+\int Dz\ \ln\int dy\ e^{\tilde{\mathcal{H}}},
    \end{align}
\end{subequations}
where 
\begin{equation}
    \begin{aligned}
        \tilde{\mathcal{H}}=&-\left(\frac{\tilde{k}}{2g}\right)y^{2}+\left(\hat{\tilde{q}}-\frac{1}{2}\hat{\tilde{Q}}\right)\phi^{2}\left(y\right)+z_{3}\sqrt{\hat{\tilde{Q}}}\phi\left(y\right)+\delta\hat{o}\phi\left(x^{*}\right)\phi\left(y\right)+\hat{p}\left(y-x^{*}\right)^{2}\\
        &\ +\tilde{k}\left(z_{2}\frac{o}{\sqrt{q}}+z_{4}\sqrt{\tilde{Q}-\frac{o^{2}}{q}}\right)y,
    \end{aligned}
\end{equation}
where $\tilde{k}=g\tilde{\beta}/\tilde{\sigma}^{2}$ and the stationary solution $\hat{O}=0$ is used. 

The remaining self-consistent SDEs are thus derived as follows,
\begin{subequations}\label{Eq. SDEs q_tilde}
    \begin{align}
        &\tilde{q}=[\langle\phi^{2}\left(y\right)\rangle],\\
        &\tilde{Q}=[\langle\phi\left(y\right)\rangle^{2}],\\
        &o=[\langle \phi\left(x^{*}\right)\phi\left(y\right)\rangle],\\
        &\hat{\tilde{q}}=-\frac{1}{2}g\tilde{k}+\frac{1}{2}g^{2}\tilde{k}^{2}\tilde{Q}+\frac{1}{2}\tilde{k}^{2}\left(1-2g\tilde{k}\tilde{Q}\right)[\langle y^{2}\rangle]+g\tilde{k}^{3}\tilde{Q}[\langle y\rangle^{2}],\\
        &\hat{\tilde{Q}}=g^{2}\tilde{k}^{2}\tilde{Q}-2g\tilde{k}^{3}\tilde{Q}[\langle y^{2}\rangle]+\tilde{k}^{2}\left(1+2g\tilde{k}\tilde{Q}\right)[\langle y\rangle^{2}],\\
        &\delta\hat{o}=\tilde{k}^{2}o\left(\frac{1}{q}-1\right)\left([\langle y^{2}\rangle]-[\langle y\rangle^{2}]\right),\\
        &[\langle\left(y-x^{*}\right)^{2}\rangle]=d^{2},
    \end{align}
\end{subequations}
where $\langle\cdot\rangle=\int dy\ e^{\tilde{\mathcal{H}}}\cdot/\int dy\ e^{\tilde{\mathcal{H}}}$.

Finally, the local entropy can be expressed as
\begin{equation}\label{LEsm}
    s={\rm ext}\{ \tilde{\mathcal{G}}_{0}+\tilde{\mathcal{G}}\},
\end{equation}
where ${\rm ext}$ indicates the local entropy depends on the converged solution of the above SDEs, which makes the physical action stationary.
The slope of the entropy with respect to $d$ is given by
\begin{equation}\label{slope}
    \frac{\partial s}{\partial d}=-2\hat{p}d.
\end{equation}
The geometry information of the high dimensional dynamics can be retrieved from these two equations [Eq.~\eqref{LEsm} and Eq.~\eqref{slope}].

\section{Numerical details of solving saddle-point equations and finding slow-point of dynamics}
\subsection{Procedure of solving SDEs}
To solve Eq.~\eqref{Eq. SDEs q}, we have to identify the minimum point $x^{*}$, which is obtained by using the golden section search method. The Gaussian integral is calculated by using the Monte Carlo method~\cite{Huang-2022}. To solve Eq.~\eqref{Eq. SDEs q_tilde}, one fixes the value of $d$ and then searches for a compatible $\hat{p}$ using the secant method~\cite{Recip-2001}. In addition, the integral $\int dy\ e^{\tilde{\mathcal{H}}}$ can be transformed to the Gaussian integral as follows,
\begin{equation}
    \int dy\ e^{\tilde{\mathcal{H}}}=\sqrt{2\pi}\sigma'\int Dy'\ e^{\tilde{\mathcal{H}}'},
\end{equation} 
where $y=\sigma'y',\ \sigma'=\left(\tilde{k}/g\right)^{-1/2}$, and 
\begin{equation}
    \tilde{\mathcal{H}}'=\left(\hat{\tilde{q}}-\frac{1}{2}\hat{\tilde{Q}}\right)\phi^{2}\left(y\right)+z_{3}\sqrt{\hat{\tilde{Q}}}\phi\left(y\right)+\delta\hat{o}\phi\left(x^{*}\right)\phi\left(y\right)+\hat{p}\left(y-x^{*}\right)^{2}+\tilde{k}\left(z_{2}\frac{o}{\sqrt{q}}+z_{4}\sqrt{\tilde{Q}-\frac{o^{2}}{q}}\right)y.
\end{equation}
A damping term would be helpful to speed up the convergence. The pseudocode is presented in Alg.~\ref{Alg. SDE Solver}. The code will be released upon formal publication of the paper~\cite{WSS-2024}.

\begin{algorithm}[H]
    \caption{SDE Solver}\label{Alg. SDE Solver}
    \begin{algorithmic}[1]
    \Require $g$, $d$, initial values of $\mathcal{O},\tilde{\mathcal{O}}$ and a damping factor $\alpha$
    \Ensure convergent values of $\mathcal{O},\tilde{\mathcal{O}}$
    \Repeat 
        \State generate Gaussian samples $z_{1},z_{2}$
        \State find $x^{*}$ by the golden section search [see Eq.~\eqref{Hfun}]
        \State calculate the average $[\cdot]$
        \State $\mathcal{O}_{t+1}\leftarrow\alpha\mathcal{O}_{t}+\left(1-\alpha\right)f\left(\mathcal{O}_{t}\right)$, where $f\left(\mathcal{O}\right)$ is the right hand side of Eq.~\eqref{Eq. SDEs q}
    \Until convergence
    \State Retain $q,x^{*},z_{2}$ for further computation
    \Repeat
        \State generate Gaussian samples $z_{3},z_{4},y'$
        \State search for $\hat{p}$ by the secant method
        \State calculate the double average $[\langle\cdot\rangle]$ in Eq.~\eqref{Eq. SDEs q_tilde}
        \State $\tilde{\mathcal{O}}_{t+1}\leftarrow\alpha\tilde{\mathcal{O}}_{t}+\left(1-\alpha\right)\tilde{f}\left(\tilde{\mathcal{O}}_{t}\right)$, where $\tilde{f}\left(\tilde{\mathcal{O}}\right)$ is the right hand side of Eq.~\eqref{Eq. SDEs q_tilde}
    \Until convergence
    \end{algorithmic}
\end{algorithm}

\subsection{Fixed-point finder}
The Levenberg-Marquart method is employed to (locally) minimize the kinetic energy~\cite{NO-2006}, and those solutions with zero velocity (actually the $\ell_2$ norm velocity is required to be below $\varepsilon$) are counted as the candidates of the fixed points $\bx_{c}$. Starting from a high number of initial points, we find that some fixed points are found repeatedly in the candidate list. If $\bx_{c}$ is different from all other candidates, it is recognized as a unique fixed point $\bx_{u}$. The fixed-point finder is terminated when the total number of the candidates $n_{c}$ is much larger than the total number of the unique points $n_{u}$, i.e. a saturation is encountered~\cite{Helias-2022}. In the large-$N$ limit, the fixed point is isolated, since if a small deviation $\{\delta_i\}$ is allowed, the deviation must obey 
the following matrix equation: $\boldsymbol{\delta}=\tilde{\mathbf{J}}\boldsymbol{\delta}$, where $\tilde{\mathbf{J}}_{ij}=J_{ij}\phi'(x_{c,j})$. This matrix equation can not be solved unless $\delta_j=0\ \forall j$ in our current setting. The pseudocode is presented in Alg.~\ref{Alg. Fixed Point Finder}.
\begin{algorithm}[H]
    \caption{Fixed Point Finder}\label{Alg. Fixed Point Finder}
    \begin{algorithmic}[1]
    \Require $\mathbf{J}$
    \Ensure $\mathbb{U}\equiv\{\mathbf{x}_{u}\}$
    \Repeat 
        \State draw the initial value $\mathbf{x}_{0}$ from a normal distribution
        \State find $\mathbf{x}_c$ by the Levenberg-Marquart method        
        \If{$\dot{\mathbf{x}}_{c}=0$}
            \State $n_{c}\leftarrow n_c+1$
            \If{$\mathbf{x}_{c}$ is unique}
                \State $\mathbb{U}\leftarrow \mathbf{x}_{c}$
                \State $n_{u}\leftarrow n_u+1$
            \EndIf
        \EndIf
    \Until $n_{u}\ll n_{c}$
    \end{algorithmic}
\end{algorithm}

\begin{figure}
    \centering
        \includegraphics[scale=0.5]{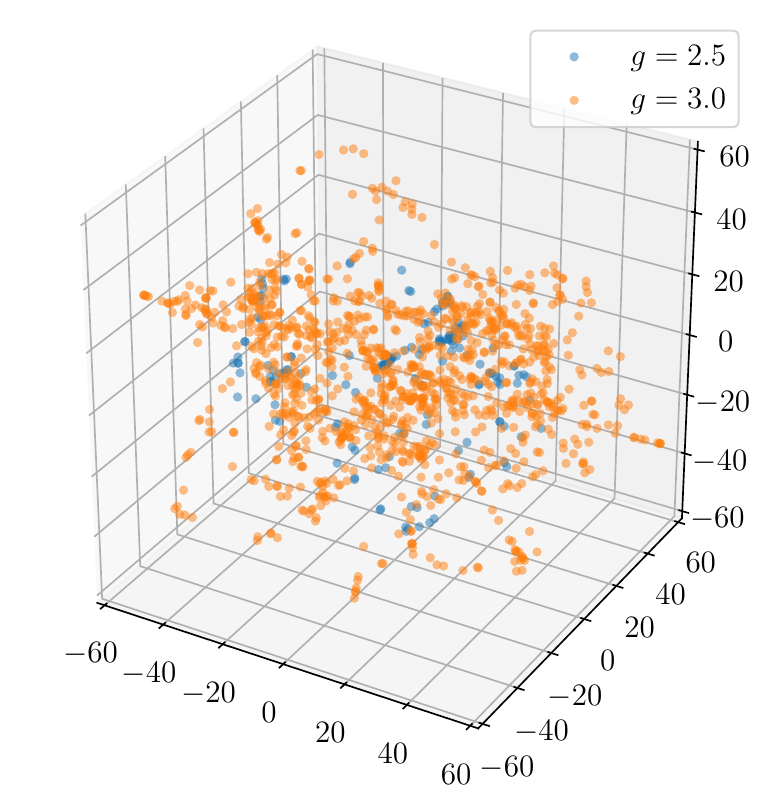}
  \caption{Low-dimensional representation of fixed points for network dynamics with $N=100$ and different values of $g$. Visualization is carried out via the nonlinear dimensionality reduction method Isomap~\cite{Sci-2000} (see details in the SM).  Fixed points are found using the Levenberg-Marquart method~\cite{NO-2006} (see SM for details) with the speed threshold $\varepsilon=10^{-6}$ .}\label{Num}
\end{figure}

In Fig.~\ref{Num}, the nonlinear dimensionality reduction method---isometric feature mapping (Isomap)~\cite{Sci-2000} is used, and a sketch of the algorithm is given in Table \ref{Table: Isomap}.  The coordinate vector $\mathbf{y}_i$ in a data set $Y$ is chosen to minimize the loss function $\|\tau(\mathbf{D}^{G})-\tau(\mathbf{D}^{Y})\|_{\ell_{2}}$, where $\|\mathbf{A}\|_{\ell_{2}}=\sqrt{\sum_{ij}A_{ij}^{2}}$, and $\mathbf{D}^{Y,G}$ is the matrix of Euclidean (for the superscript $Y$) or geodesic (for the superscript $G$)
distances for points in $Y$. The operator $\tau$ is defined by $\tau(\mathbf{D})=-\mathbf{HSH}/2$, where $\mathbf{S}_{ij}=\mathbf{D}_{ij}^2$, and $\mathbf{H}_{ij}=\delta_{ij}-1/n$ ($n$ denotes the number of high-dimensional points in $Y$). This minimization finds an embedding that preserves the interpoint distances as much as possible.
\begin{table}[H]
    \centering
    \caption{Isomap steps.}\label{Table: Isomap}
    \begin{ruledtabular}
    \begin{tabular}{lll}
    1 & Neighborhood graph construction & Connect points $i$ and $j$ if their distance $D_{ij}^{Y}$ is shorter than $\epsilon$.\\
    2 & Geodesic distance $D_{ij}^{G}$ & Initialize $D^{G}_{ij}=D_{ij}^{Y}$ if $i,j$ are connected; $D^{G}_{ij}=\infty$ otherwise.\\
    &&  Update $D^{G}_{ij}=\min\{D_{ij}^{G},D_{ik}^{G}+D_{kj}^{G}\},\ k=1,2,\cdots,n$.\\
    3 & $d$-dimensional embedding & The dimension-reduced coordinates $Y=(\mathbf{E}^{d})^\top(\Lambda^{d})^{1/2},$\\
    && where $\Lambda^{d}$ is the diagonal matrix of the $d$ largest eigenvalues of $\tau(\mathbf{D}^{G})$,\\
    &&  and the associated eigenvectors form $\mathbf{E}^d$.\\
    \end{tabular}
    \end{ruledtabular}
\end{table}

\end{document}